# Imprinting of skyrmions and bimerons in an antiferromagnet

Coline Thevenard,[1,*] Miina Leiviskä,[1] Richard F. L. Evans,[2,†] Daria Gusakova,[1] Vincent Baltz[1,‡]

[1]*Univ. Grenoble Alpes, CNRS, CEA, Grenoble INP, IRIG-SPINTEC, F-38000 Grenoble, France*

[2]*School of Physics, Engineering and Technology, University of York, York YO10 5DD, United Kingdom*

**coline.thevenard@cea.fr; †richard.evans@york.ac.uk; ‡vincent.baltz@cea.fr*

## Abstract

Topologically protected magnetic states in condensed matter physics, particularly antiferromagnetic (AFM) skyrmions (Sks) and bimerons (Bms), offer promising prospects for terahertz dynamics and sustained current-induced motion, thanks to their compensating multiple sub-lattice structure. However, nucleating AFM Sks and Bms is challenging due to the lack of net magnetization. Previous attempts to imprint pre-defined Sks and Bms in a ferromagnet (FM) and transfer them to an AFM using interfacial exchange bias in FM/AFM heterostructures have been hindered by complex multilayers with discontinuities, polycrystallinity, or multipartite chiral AFMs. Employing atomistic spin simulations, we demonstrate the viability of texture imprinting for nucleating Sks and Bms in AFMs, using a prototypical bipartite AFM layer in a multilayer structure free from discontinuities. Such imprinting is a crucial step towards understanding the static and dynamic properties of natural AFM textures and their unique properties.

## I. Introduction

The interest in topologically protected states in condensed matter physics stems from their robustness against small perturbations and their significant impact on the physical properties of the system. This impact is closely tied to the system's symmetry, which for example enables a non-zero geometric phase for electrons interacting with the spin structure. While topological objects in uncompensated magnets, like ferromagnets (FMs), have been extensively studied due to their fundamental significance and potential applications [1,2], the presence of multiple spin sub-lattices in compensated magnets with zero net magnetization, like antiferromagnets (AFMs) and altermagnets opens new avenues for topological phenomena and spintronic applications [3–8]. Specifically, real-space topological states, such as AFM skyrmions (Sks) and bimerons (Bms) - in-plane topological homeomorphic counterparts of Sks - exhibit intriguing properties. Due to anti-parallel interatomic exchange interactions, AFM Sks and Bms possess zero net magnetization and topological charge, ensuring robustness against external fields and vanishing net topological charge and Hall effects [9–11]. Moreover, they are predicted to induce unique transport properties, including a non-vanishing topological spin Hall effect [12,13] and a longitudinal Sk velocity unconstrained by the Sk Hall effect, exceeding that of FM Sks due to the inherent THz dynamics in AFMs [14,15]. These properties make AFM Sks and Bms not only fascinating subjects for studies on topology but also promising candidates for ultra-dense, ultra-fast, low-power spintronic devices. However, the main challenge in accessing these advantageous properties is the difficulty in experimentally nucleating AFM Sks and Bms due to the lack of net magnetization. Various nucleation techniques have been proposed, including field-assisted methods, magneto-electric effects and ultrafast laser pulses [7]. An alternative approach involves pre-defining Sks and Bms in a FM and subsequently imprinting them in the AFM using interfacial exchange bias in FM/AFM heterostructures, facilitated by appropriate thermal and magnetic field cycles. This method has successfully transferred spin textures in AFMs, such as vortices, domain walls and non-topologically protected bubbles [7,16–20]. For Sks and Bms, both simulations and experiments have attempted to validate this imprinting method, but complex multilayers with discontinuities, polycrystallinity, or multipartite chiral AFMs have hindered accurate calculations of topological parameters [7,19,20]. In this study, we employ atomistic simulations to demonstrate the viability of texture imprinting for nucleating Sks and Bms in AFMs. Using a prototypical bipartite AFM layer in a multilayer structure free from discontinuities, we quantitatively confirm the imprinting of actual Sks and Bms in an AFM, starting from pre-defined FM Sks and Bms [Fig. 1].

## II. Atomistic simulation system

The numerical study focuses on a single-crystal AFM (*n*)/NM (1)/FM (3)/HM (1 ML) system [Fig. 1(a)], where the number of monolayers (MLs) is indicated in parenthesis, with *n* ranging from 11 to 28 MLs.

The crystal structure is consistently set to that of two interpenetrating single cubic lattices and equivalent to rocksalt. The lattice parameter is of 4.17 Å, that of the NiO AFM [21,22]. The area of the simulation system is 100×100 nm$^2$. The crystal is oriented along the [001] direction, taken as the z-axis and features a continuous structure. Periodic boundary conditions are applied in the x ([100] direction) and y ([010] direction), to approach an infinite system. The roles of the non-magnetic (NM) layer in inducing magnetic frustration for exchange bias coupling and the heavy metal (HM) layer in promoting perpendicular anisotropy and spin canting to stimulate spin textures formation [18,19] will be detailed later in the text.

Simulations are performed using the VAMPIRE atomistic simulation software [23] on the ARCHER2 supercomputer, each utilizing 1024 processors across 8 physical nodes. The large-scale simulation involves $10^6$ time steps and employs the adaptative Monte Carlo algorithm to minimize the energy of up to $7.6\times10^6$ spins [24].

The interactions within the system are governed by the spin Hamiltonian [25]

$$\mathcal{H} = -\sum_{i<j} J_{ij}\left(\mathbf{S}_i \cdot \mathbf{S}_j\right) - \sum_{i<j} \mathbf{D}_{ij} \cdot \left(\mathbf{S}_i \times \mathbf{S}_j\right) - \sum_{i<j} k_{N,ij}\left(\mathbf{S}_i \cdot \mathbf{e}_{ij}\right)^2 - \sum_{i} k_{u,i}(\mathbf{S}_i \cdot \mathbf{e}_i)^2 - \sum_{i} \mu_{s,i}\, \mathbf{S}_i \cdot \mathbf{B}_{ext}$$

where the indices i and j refer to the position of the atomic sites occupied by spins of vector **S** and magnetic moment $\mu_s$. The Heisenberg exchange interaction, parametrized by the constant $J_{ij}$, is represented by the first term. The second term accounts for the Dzyaloshinskii-Moriya interaction (DMI), described by the vector $\mathbf{D}_{ij} = D_{ij}\left(\hat{\mathbf{r}}_{ik} \times \hat{\mathbf{r}}_{jk}\right)$, where $D_{ij}$ is the DMI strength and $\hat{\mathbf{r}}_{ik,jk}$ the unit vector joining the magnetic atoms at sites i and j and the non-magnetic atom at site k. The third term introduces a radial spin rotation with respect to a specific nearest neighbor, relating to the Néel surface anisotropy characterized by the constant $k_{N,ij}$. The fourth term introduces a uniaxial anisotropy of constant $k_{u,i}$, along the axis of unit vector $\mathbf{e}_i$ and the last term is the Zeeman interaction, which couples the spins to an external induction $\mathbf{B}_{ext}$.

The magnetic properties of the HM/FM bilayer are based on those of Pt/Co which have been optimized to nucleate Sks in the FM [18,26]. Specifically, the interfacial perpendicular magnetic anisotropy is $k_{N,HM-FM}$ = 2×10$^{-24}$ J/atom and the DMI is $D_{HM-FM}$ = 10$^{-21}$ J/link. The FM has an atomic magnetic moment $\mu_{s,FM}$ = 1.37 $\mu_B$, an exchange constant $J_{FM}$ = 13.3×10$^{-21}$ J/link (which corresponds to the Curie temperature $T_C$ = 1385 K, from $J_{FM} = \frac{3k_BT_C}{\epsilon z}$ [23] with $\epsilon = 0.719$ [22] and $z = 6$ here) and (i) a uniaxial anisotropy $k_{u,FM}$ = 10$^{-24}$ J/atom for the Sk case [Fig. 1(b)] or (ii) a hard-axis anisotropy $k_{u,FM}$ = -10$^{-23}$ J/atom for the Bm case [Fig. 1(c)]. In its ground state, the AFM is composed of two collinear spin sublattices AFMa and AFMb, hosting an atomic magnetic moment $\mu_{s,AFM}$ = 2 $\mu_B$, that of NiO [21], coupled antiparallel via a negative exchange constant $J_{AFM}$ = -5.04×10$^{-21}$ J/link (which corresponds to the Néel temperature $T_N$ = 525 K of NiO [27]), and subject to: (i) a uniaxial anisotropy along the z-axis $k_{u,AFM}$ = $10^{-25}$ J/atom for the Sk case [Fig. 1(b)], or (ii) a hard-axis anisotropy $k_{u,AFM}$ = $-10^{-24}$ J/atom for the Bm case [Fig. 1(c)]. The FM and the AFM are coupled via an interfacial exchange interaction $J_{FM-AFM}$ = $10^{-21}$ J/link, unless specified otherwise. We note that conversion to SI units is provided in Table S1 in the SM. To imprint spin textures in the AFM, exchange bias coupling requires a net interfacial magnetic moment [25]. A NM monolayer mixed with FM and populated with 25% of substituted FM atoms, unless specified otherwise, generates a spatial distribution of magnetic exchange along the interface. This distribution results in a net interfacial magnetic moment within the AFM, providing a reasonable exchange bias, while preserving a flat interface, which is essential for further calculating the topology of the imprinted textures in the AFM. The influence of FM substitution on texture imprinting is detailed later in the text, with the example of Sk imprinting.

## III. Imprinting of AFM spin textures

The nucleation and stabilization processes of AFM Sks are depicted in Fig. 2. They require three successive simulation steps [7,16–20,26]. First, Sks are formed in the FM layer only, at 600 K under an applied field of 0.5 T along the z-axis. In this step, the FM is magnetically ordered [Figs. 2(a) and S1 of the Supplemental Materials (SM)] while the AFM remains disordered, as 600 K is below the FM's Curie temperature $T_C$ = 1385 K and above the AFM's Néel temperature $T_N$ = 525 K [Figs. 2(b,c) and S1]. The effect of the field strength and DMI in favoring the formation of Sks in the FM over maze domains or full saturation is shown in Fig. S2. Next, AFM spin textures are imprinted via exchange coupling at the FM/AFM interface. The system is cooled from 600 to 0 K, in the same applied field of 0.5 T. This thermal process leads to the collapse of one Sk and the stabilization of the other in the FM [Fig. 2(d,g)]. Upon crossing $T_N$, the AFM orders and interfacial coupling is triggered, setting the Sk in the AFM [Fig. 2(e,f)]. The resulting AFM texture consists of two interlocked FM Sks with opposite polarity [Fig. 2(h,i)]. We note that using 0 K avoids thermal noise, facilitating the visual identification and quantification of the spin textures' topological nature, as will be detailed later in the text. Finally, the applied field is removed to reach remanence. This step widens the FM and AFM Sk's radius and wall width [Fig. 2(j-l)], as expected from the field-dependent energy landscape of FM Sks [28].

We note that, due to the high computational cost of the simulations, the dipole-dipole interactions were not included for the calculations. Here, the dipole fields, implemented using the tensor approach described in [29] are shown to have a quantitative effect acting as an effective in-plane anisotropy leading to the formation of Bms. Increasing the perpendicular anisotropy compensates for this effect, leading to the reappearance of Sks (Fig. S3).

The strength of the interfacial coupling plays an important role in the formation of the Sks, as shown in Fig. 3. For weak coupling between the FM and AFM, no Sk is imprinted [see also Fig. S4], but increasing the exchange coupling increases the overall effective exchange in the system, leading to a decrease in the FM Sk size. For a critical strength of $5\times10^{-22}$ J/link, Sks do not form at all, as the interfacial exchange is insufficient to sustain them in the AFM and propagate the DMI into the AFM layer [see also Fig. S5]. For higher exchange coupling the FM and AFM layers become strongly coupled, leading to direct imprinting of Sks from the FM to the AFM. As the interfacial exchange is increased further, there is a small decrease in the Sk size due to the overall increase in exchange in the system, but this is a relatively weak effect.

We next discuss the influence of FM substitution in the NM interfacial layer, on texture imprinting. For the case of 0% FM atoms, no exchange coupling occurs as there are no exchange bonds between the FM and the AFM. Sks form in the FM but they are naturally not replicated in the AFM, which then maintains two sublattices with opposite out-of-plane spins. We note that the Sk size of 13.9 nm matches the case when $J_{FM-AFM} = 0$ J/link (Fig. 3(a)), showing a radius of 13.2 nm, in line with the analytical calculation that can be made in this specific case: 15.3 nm from $R = \pi D_{HM-FM}\sqrt{\frac{J_{FM}}{16 J_{FM} k_{eff}^2 - \pi^2 D_{HM-FM}^2 k_{eff}}}$ [28] where $k_{eff} = k_{u,FM} - \frac{2\times k_{N,HM-FM}}{t_{FM}}$ with $t_{FM}$ the thickness of the FM. We further note that the difference in FM Sk density between Fig. 4 (0% FM atoms leading to virtually no coupling) and Fig. 3 ($J_{FM-AFM} = 0$ with 25% FM bonds, also leading to no coupling), is likely due to Sk pinning at the mixed interface. Substituting some NM atoms, 25% 50% and 75%, creates a sufficient net interfacial magnetic moment, enabling texture imprinting in the AFM. However, the FM Sk size (and consequently the AFM Sk size) decreases as the percentage of FM atoms increases due to stronger exchange bias, resulting in the FM Sk experiencing a larger effective external field [28]. We anticipate similar effects for Bms. With a full FM layer at the interface (100 %), exchange bias becomes too large, preventing Sk formation in the FM. Due to frustration (the two AFM sublattices cannot simultaneously achieve parallel coupling with the FM and maintain antiparallel coupling between themselves), the AFM undergoes a spin-flop, with antiparallel spins lying in-plane.

## IV. Tubular skyrmion

The assignment of the given topological class to the Sks is obtained by calculating their topological charge Q [Fig. 5(a)] [1,2,7]. For a two-dimensional discrete lattice of spins (i.e. for each ML in the system) Q is a sum of local densities q calculated for a square cell [insert of Fig. 5(a)] [30,31]:

$$q(x^*) = \frac{1}{4\pi}\sum_{(i,j,k)\in\{(1,2,3),(1,3,4)\}} 2\tan^{-1}\left(\frac{\mathbf{s}_i\cdot(\mathbf{s}_j\times\mathbf{s}_k)}{1+\mathbf{s}_i\cdot\mathbf{s}_j+\mathbf{s}_i\cdot\mathbf{s}_k+\mathbf{s}_j\cdot\mathbf{s}_k}\right).$$

Figure 5 demonstrates the topological nature of the nucleated Sks in the FM below 529 K, with a topological charge Q = 2, which reduces to 1 upon annihilation of one FM Sk at 359 K. It also demonstrates the topological nature of the imprinted Sk in the AFM below 207 K with Q = +(−) 1 for the AFMb (AFMa) sublattices. The values were confirmed throughout the depth of the AFM. At higher temperatures, thermal fluctuations can cause neighboring spins to deviate significantly from each other resulting in deviation of Q from an integer value [30] [Fig. 5(b-m)]. Note that the Sk imprinted in the AFM consists of two intercalated FM Sks with mutually reversed spins [Fig. 6(a)] resulting in an AFM Sk with a net topological charge of zero [32].

To investigate the intricate topology of the Sks formed with more accuracy, the usual three key parameters are introduced and calculated: polarity, vorticity and helicity [1,2,7]. Polarity p quantifies the out-of-plane magnetization density of the Sk core, vorticity m represents the magnetization density winding number, and helicity γ describes the angular phase shift between the spatial magnetization rotation and the positional winding. We note that Q = pm. In the final state [Fig. 2(j-l)], (p, m, γ) = (1, 1, π) for the FM layer and (-1, 1, 0) and (1, 1, π) for the AFMa and AFMb sublattices, respectively. These values indicate that the initial Sk in the FM is of the Néel type, consistent with the DMI strength of $10^{-21}$ J/link which exceeds the critical strength for a Bloch to Néel domain wall transition $D_c = 4\mu_0 M_s^2 t \ln 2/(2\pi^2)$ = $10^{-23}$ J/link [33]. The Sk imprinted in the AFM sublattices naturally replicates the morphology and topology (in terms of p, m and γ) of the Sks from which it originates in the FM layer.

Due to the regularity of the atomic lattice and continuity at the interfaces, the FM (AFM) Sk in its final state [Fig. 2(j-l)] forms a disc with in-plane spins describing a circle of radius R = 7.8 (7.1) nm and wall width w = 8.5 (9.7) nm, estimated by fitting $s_z(r) = \cos(2\tan^{-1}(\sinh(R/w)\,/\,\sinh(r/w)))$ [34]. The close values for both materials clearly indicate that the arrangement of the AFM spins follows that of the FM spins.

Next, we investigate how the topological texture propagates from the interface through the depth of the AFM and how temperature affects this propagation. To quantify the depth-dependent arrangement of each AFM sublattice spins relative to the FM spins, we employ the partial spin-spin correlation function [Fig. 6(b)]:

$$C(\Delta z) = \frac{1}{N}\sum_{(i,j)} s_{z,i}\, s_{z,j}(\Delta z)$$

where Δz is the distance to the FM/AFM interface and N is the number of spin pairs formed by the FM spin i and AFM j of the same in-plane coordinates (x,y) [19]. This quantitative method enhances the bare observation of cross-sections [Fig. 6(c)]. The sign of C is naturally opposite when considering the two AFM sublattices of opposite spins.

Starting from fully anticorrelated spins ($C \approx 0$) at 600 K, when the AFM is disordered, the system gradually reaches full correlation ($|C| \approx 1$) at 0 K, across the entire AFM layer. This indicates that the replication is effective throughout the 5.8 nm-thick AFM layer, which corresponds to the maximum penetration depth allowed for this set of parameters [Fig. S6]. We note however that, when the correlation remains constant in the core of the AFM layer, it deteriorates in the few MLs at the FM/AFM interface due to the loss of surface exchange bonds and a local decrease in AFM surface order. Similarly, the correlation also deteriorates in the last few MLs away from the free AFM edge, due to missing magnetic bonds. In other words, close to the edge, the decorrelation from the FM prevents the rearrangement of the AFM spins. We also consider the different energy terms during formation of the Sks, and see no significant difference of energy for the system with and without Sks [Fig. S7]. Unlike micromagnetic simulations, the total energy is dominated by exchange and so an analysis of the energy does not provide significant insights into the imprinting process.

## V. Tubular bimeron

To test the universality of the replication, we next simulate the same heterostructure under identical energetic conditions, with the sole modification being the change of anisotropy from out-of-plane to in-plane. Specifically, the FM constant $k_{u,FM}$ = -$10^{-23}$ J/atom and the AFM constant $k_{u,AFM}$ = - $10^{-24}$ J/atom now confine the magnetization to the xy-plane. This allows us to nucleate an in-plane Sk, or Bm in the FM layer and to imprint it throughout the thickness of the AFM layer [Figs. 1(c) and 7], using the thermal and field procedure [Figs. S8 and S9] described earlier in the text. In the final state, the axis of the Bm, which joins the centers of the two merons is angled by -3.3° in the xy-plane with respect to the x-axis. This in-plane orientation is a consequence of using z as the hard axis, allowing the Bm axis to lie anywhere within the isotropic xy-plane. Additionally, the axis is slightly titled out-of-plane by 1.5° ascribed to magnetic frustration at the FM/AFM interface, combined with the residual interfacial anisotropy $k_N$. Moreover, by analogy with a Sk, the topological parameters of the Bm are (p, m, $\gamma$) = (1, 1, $\pi$) for the FM layer, and (-1, 1, 0) and (1, 1, $\pi$) for the AFMa and AFMb sublattices, respectively. These parameters were calculated after applying a spin rotation operation in spin space to align the Bm axis with the z-axis. The invariance of the topological parameters, along with an identical topological charge Q = 1 for the FM layer and AFMb sublattice and -1 for the AFMa sublattice, confirms that the texture we imprinted in the AFM layer is a Bm, a quasiparticle homeomorphic to the Sk imprinted in the first part of this work [9,35].

## VI. Conclusion

In conclusion, we have theoretically validated the feasibility of texture imprinting for the nucleation of actual skyrmions and bimerons in a compensated magnet with zero net magnetization, employing predefined skyrmions and bimerons in an adjacent exchange-coupled ferromagnet as a template. Our study involves engineering a prototypical bipartite antiferromagnetic layer within a multilayer structure, ensuring the presence of the necessary Dzyaloshinskii-Moriya interaction (DMI), perpendicular anisotropy, and interfacial magnetic frustrations, all while avoiding discontinuities. This approach has enabled us to calculate the topological parameters of the imprinted spin textures, providing quantitative data that was previously inaccessible due to the complexities of multilayers with discontinuities, polycrystallinity, or multipartite chiral antiferromagnets. We have successfully tracked the formation of the topological objects throughout the thermal procedure, first nucleating skyrmions and bimerons in the ferromagnet, and then replicating them in the antiferromagnet after crossing the Néel temperature. The replicated skyrmions and bimerons were found to extend beyond the interface, spanning the core of the antiferromagnet. This work addresses the challenge of nucleating quantitatively proven topological magnetic textures in compensated magnets with zero net magnetization. Its primary interest is fundamentally driven as it paves the way for extending the study of topological spin textures beyond ferromagnets, into materials with intriguing electrodynamic properties, such as antiferromagnets and altermagnets. Specifically, because the method is independent of the electronic nature of the materials used, one could envisage an insulating ferromagnet and a metallic antiferromagnet, potentially in contact with a heavy metal, for the fundamental study of the dynamics of skyrmion propagation in antiferromagnets under current by spin torques, and the reciprocal effects concerning the influence of an antiferromagnetic skyrmion on electron dynamics. Given that the dynamics of antiferromagnets are in the THz regime, from an applicative perspective, one could imagine nucleating a skyrmion in a ferromagnet and then leveraging the THz dynamics of antiferromagnet, unconstrained by the skyrmion Hall effect [14,15] to propagate information, detecting the antiferromagnetic skyrmion via non-vanishing topological spin Hall effects [12,13] or through the inverse nucleation of a ferromagnetic skyrmion. The use of this method for applications will be limited by the speed of imprinting. However, one could envisage the use of femtosecond lasers for the combined nucleation of ferromagnetic skyrmions [36] and imprinting of antiferromagnetic skyrmions.

## Acknowledgments

This study was partially supported by the CNRS international research project (IRP) "CITRON", the U.K. EPSRC International Network for Spintronics (Grant No. EP/V007211/1), and the French National Research Agency through the France 2030 government grant PEPR-SPIN "ALTEROSPIN" ANR-24-EXPR-0002. This work used the ARCHER2 UK National Supercomputing Service.

## Data availability statement

All data are provided in full in the main text and supplementary information of this paper. Raw data supporting the findings of this study are available [ https://doi.org/10.57745/GCJWSK ].

## References

[1] A. Soumyanarayanan, N. Reyren, A. Fert, and C. Panagopoulos, Emergent phenomena induced by spin–orbit coupling at surfaces and interfaces, Nature **539**, 509 (2016).

[2] B. Göbel, I. Mertig, and O. A. Tretiakov, Beyond skyrmions: Review and perspectives of alternative magnetic quasiparticles, Phys. Rep. **895**, 1 (2021).

[3] T. Jungwirth, X. Marti, P. Wadley, and J. Wunderlich, Antiferromagnetic spintronics, Nat. Nanotechnol. **11**, 231 (2016).

[4] V. Baltz, A. Manchon, M. Tsoi, T. Moriyama, T. Ono, and Y. Tserkovnyak, Antiferromagnetic spintronics, Rev. Mod. Phys. **90**, 015005 (2018).

[5] L. Šmejkal, J. Sinova, and T. Jungwirth, Emerging Research Landscape of Altermagnetism, Phys. Rev. X **12**, 040501 (2022).

[6] V. Bonbien, F. Zhuo, A. Salimath, O. Ly, A. Abbout, and A. Manchon, Topological aspects of antiferromagnets, J. Phys. D. Appl. Phys. **55**, 103002 (2022).

[7] M. Leiviskä et al., Antiferromagnetic skyrmions in spintronics, Book Chapter, Elsevier (Eds A. Hirohata al.), hal-04390931 (2024).

[8] O. J. Amin et al., Antiferromagnetic half-skyrmions electrically generated and controlled at room temperature, Nat. Nanotechnol. **18**, 849 (2023).

[9] B. Göbel, A. Mook, J. Henk, I. Mertig, and O. A. Tretiakov, Magnetic bimerons as skyrmion analogues in in-plane magnets, Phys. Rev. B **99**, 060407 (2019).

[10] L. Shen, X. Li, J. Xia, L. Qiu, X. Zhang, O. A. Tretiakov, M. Ezawa, and Y. Zhou, Dynamics of ferromagnetic bimerons driven by spin currents and magnetic fields, Phys. Rev. B **102**, 104427 (2020).

[11] L. Shen, J. Xia, X. Zhang, M. Ezawa, O. A. Tretiakov, X. Liu, G. Zhao, and Y. Zhou, Current-Induced Dynamics and Chaos of Antiferromagnetic Bimerons, Phys. Rev. Lett. **124**, 037202 (2020).

[12] J. Barker and O. A. Tretiakov, Static and Dynamical Properties of Antiferromagnetic Skyrmions in the Presence of Applied Current and Temperature, Phys. Rev. Lett. **116**, 147203 (2016).

[13] E. G. Tveten, A. Qaiumzadeh, O. A. Tretiakov, and A. Brataas, Staggered Dynamics in Antiferromagnets by Collective Coordinates, Phys. Rev. Lett. **110**, 127208 (2013).

[14] P. M. Buhl, F. Freimuth, S. Blügel, and Y. Mokrousov, Topological spin Hall effect in antiferromagnetic skyrmions, Phys. Status Solidi - Rapid Res. Lett. **11**, 1700007 (2017).

[15] C. A. Akosa, O. A. Tretiakov, G. Tatara, and A. Manchon, Theory of the Topological Spin Hall Effect in Antiferromagnetic Skyrmions: Impact on Current-Induced Motion, Phys. Rev. Lett. **121**, 097204 (2018).

[16] G. Salazar-Alvarez et al., Direct evidence of imprinted vortex states in the antiferromagnet of exchange biased microdisks, Appl. Phys. Lett. **95**, 012510 (2009).

[17] J. Wu et al., Direct observation of imprinted antiferromagnetic vortex states in CoO/Fe/Ag(001) discs, Nat. Phys. **7**, 303 (2011).

[18] K. G. Rana et al., Imprint from ferromagnetic skyrmions in an antiferromagnet via exchange bias, Appl. Phys. Lett. **119**, 192407 (2021).

[19] M. Leiviskä, S. Jenkins, R. F. L. Evans, D. Gusakova, and V. Baltz, Local setting of spin textures in a granular antiferromagnet, Phys. Rev. B **108**, 184424 (2023).

[20] B. He et al., Creation of Room-Temperature Sub-100 nm Antiferromagnetic Skyrmions in an Antiferromagnet IrMn through Interfacial Exchange Coupling, Nano Lett. **24**, 2196 (2024).

[21] P. García-Fernández, J. C. Wojdeł, J. Íñiguez, and J. Junquera, Second-principles method for materials simulations including electron and lattice degrees of freedom, Phys. Rev. B **93**, 195137 (2016).

[22] D. Garanin, Self-consistent Gaussian approximation for classical spin systems: Thermodynamics, Phys. Rev. B - Condens. Matter Mater. Phys. **53**, 11593 (1996).

[23] R. F. L. Evans, W. J. Fan, P. Chureemart, T. A. Ostler, M. O. A. Ellis, and R. W. Chantrell, Atomistic spin model simulations of magnetic nanomaterials, J. Phys. Condens. Matter **26**, 103202 (2014).

[24] J. D. Alzate-Cardona, D. Sabogal-Suárez, R. F. L. Evans, and E. Restrepo-Parra, Optimal phase space sampling for Monte Carlo simulations of Heisenberg spin systems, J. Phys. Condens. Matter **31**, 095802 (2019).

[25] S. Jenkins, Spin Dynamics Simulations of Iridium Manganese Alloys, University of York, 2020.

[26] S. Brück, J. Sort, V. Baltz, S. Suriñach, J. S. Muñoz, B. Dieny, M. D. Baró, and J. Nogués, Exploiting Length Scales of Exchange-Bias Systems to Fully Tailor Double-Shifted Hysteresis Loops, Adv. Mater. **17**, 2978 (2005).

[27] N. Rinaldi-Montes, P. Gorria, D. Martínez-Blanco, A. B. Fuertes, I. Puente-Orench, L. Olivi, and J. A. Blanco, Size effects on the Neél temperature of antiferromagnetic NiO nanoparticles, AIP Adv. **6**, 056104 (2016).

[28] X. S. Wang, H. Y. Yuan, and X. R. Wang, A theory on skyrmion size, Commun. Phys. **1**, 31 (2018).

[29] G. J. Bowden, G. B. G. Stenning, and G. Van Der Laan, Inter and intra macro-cell model for point dipole-dipole energy calculations, J. Phys. Condens. Matter **28**, (2016).

[30] B. Berg and M. Lüscher, Definition and statistical distributions of a topological number in the lattice O(3) σ-model, Nucl. Phys. B **190**, 412 (1981).

[31] M. Böttcher, S. Heinze, S. Egorov, J. Sinova, and B. Dupé, B-T phase diagram of Pd/Fe/Ir(111) computed with parallel tempering Monte Carlo, New J. Phys. **20**, 103014 (2018).

[32] O. A. Tretiakov, D. Clarke, G. W. Chern, Y. B. Bazaliy, and O. Tchernyshyov, Dynamics of Domain Walls in Magnetic Nanostrips, Phys. Rev. Lett. **100**, 127204 (2008).

[33] A. Thiaville, S. Rohart, É. Jué, V. Cros, and A. Fert, Dynamics of Dzyaloshinskii domain walls in ultrathin magnetic films, Europhys. Lett. **100**, 57002 (2012).

[34] H. B. Braun, Fluctuations and instabilities of ferromagnetic domain-wall pairs in an external magnetic field, Phys. Rev. B **50**, 16485 (1994).

[35] Y. Shen, Q. Zhang, P. Shi, L. Du, X. Yuan, and A. V. Zayats, Optical skyrmions and other topological quasiparticles of light, Nat. Photonics **18**, 15 (2024).

[36] S. G. Je et al., Creation of Magnetic Skyrmion Bubble Lattices by Ultrafast Laser in Ultrathin Films, Nano Lett. **18**, 7362 (2018).

## Figures and captions

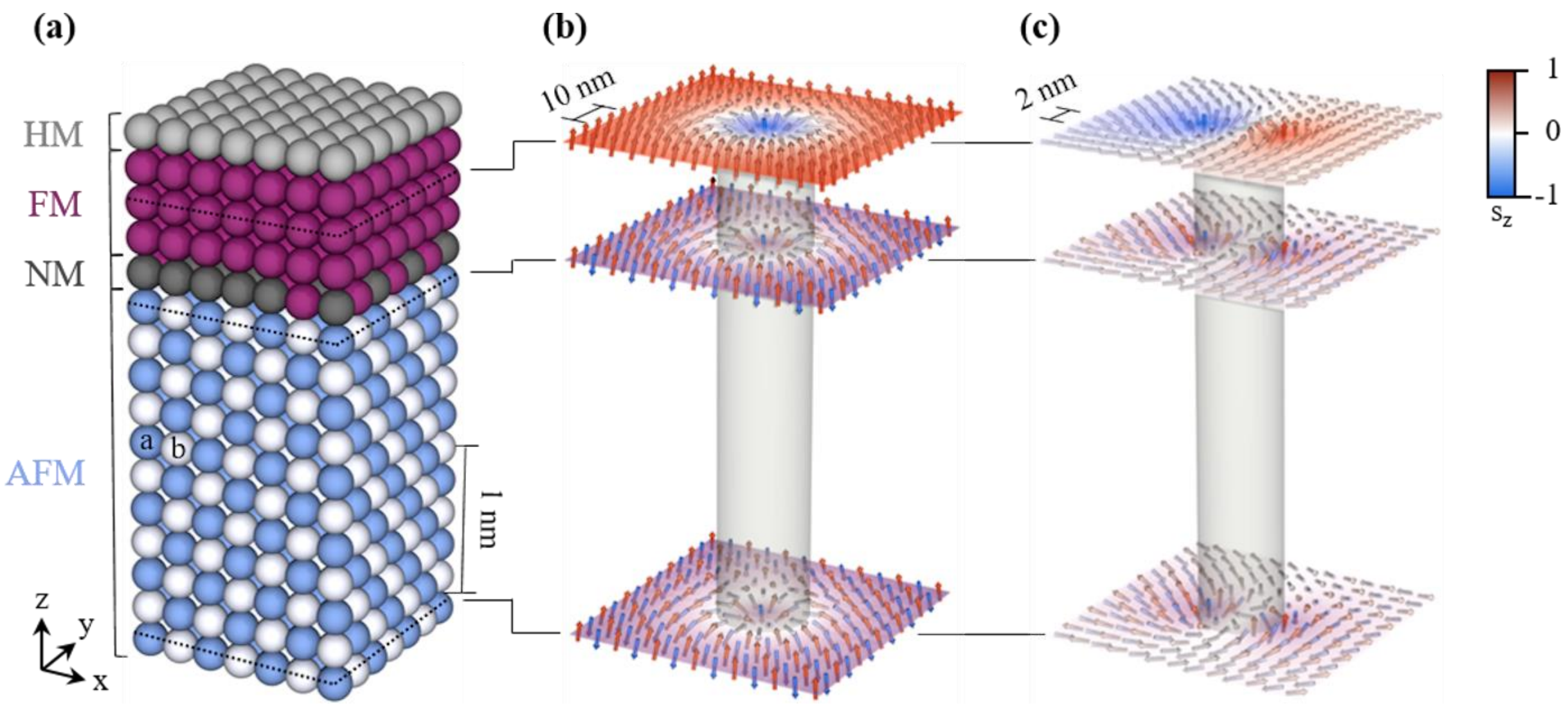

Figure 1. (a) The atomic structure of the simple cubic monocrystalline simulation system consisting of: a HM/FM bilayer for Sk nucleation, a NM interface substituted with 25% of FM atoms for coupling, and a prototypical bipartite AFM layer made of two sublattices: AFMa and AFMb. In their ground state, the spins of AFMa and AFMb are collinear and opposite. The full size of the simulation system is 100 × 100 × 3.34 $nm^3$ corresponding to 1 monolayer (ML) of HM, 3 MLs of FM, 1 ML of NM and 11 MLs of AFM. 3D views of the Sk (b) and Bm (c) tubes obtained in the FM and imprinted in the 11 MLs-thick AFM, following thermal and magnetic field cycling procedures. The simulated spins oriented in the xy- (b) yz- (c) plane define the contour of the tube.

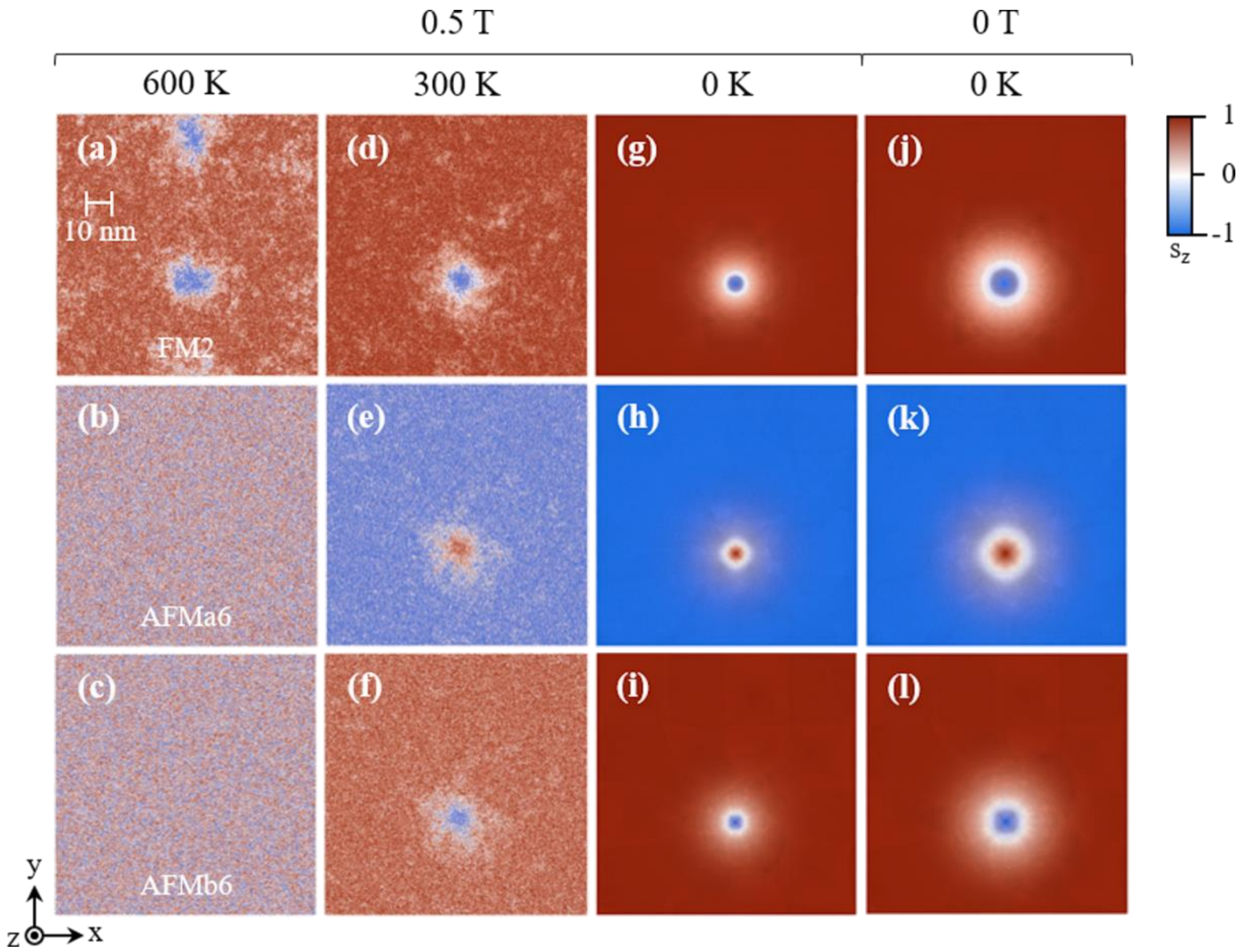

Figure 2. Top view cross section snapshots of the evolution of the spin textures in the FM (first row, 2nd ML from the FM/AFM interface), and the AFM in the core (second row: AFMa sublattice, third row: AFMb sublattice, 6th ML from the FM/AFM interface). First column: field-induced nucleation in the FM, below the Curie temperature ($T_C$) of the FM and above the Néel temperature ($T_N$) of the AFM [Fig. S1]. Second and third columns: imprinting and stabilization of Sks in the AFM, via cooling across $T_N$. Fourth column: remanence state.

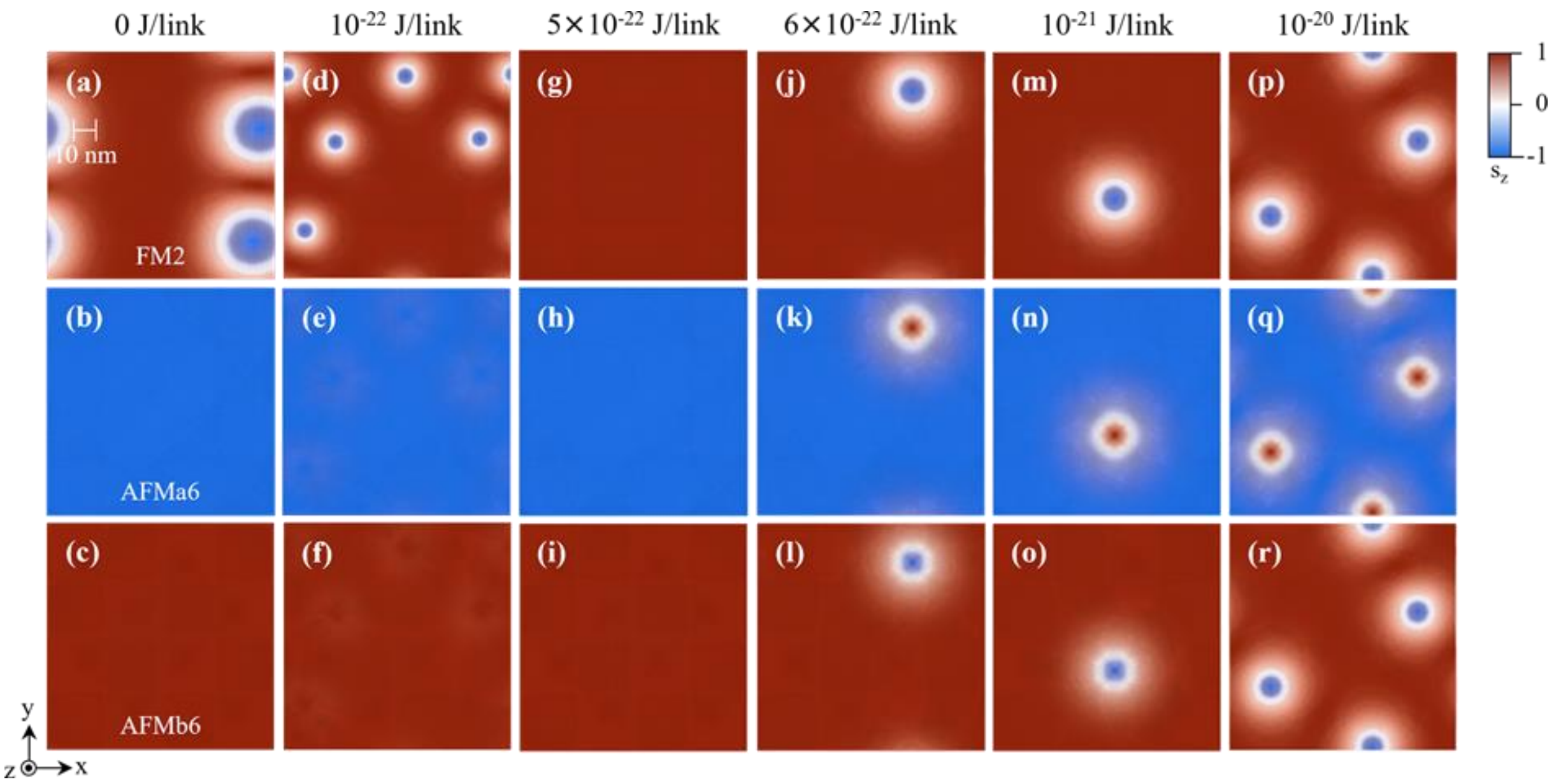

Figure 3. Top view cross section snapshots of the evolution of the final state of the Sk in the FM (first row, 2nd ML from the interface), and the AFM in the core (second row: AFMa, third row: AFMb, 6th ML from the interface) under the effect of the FM-AFM exchange interaction, $J_{FM-AFM}$.

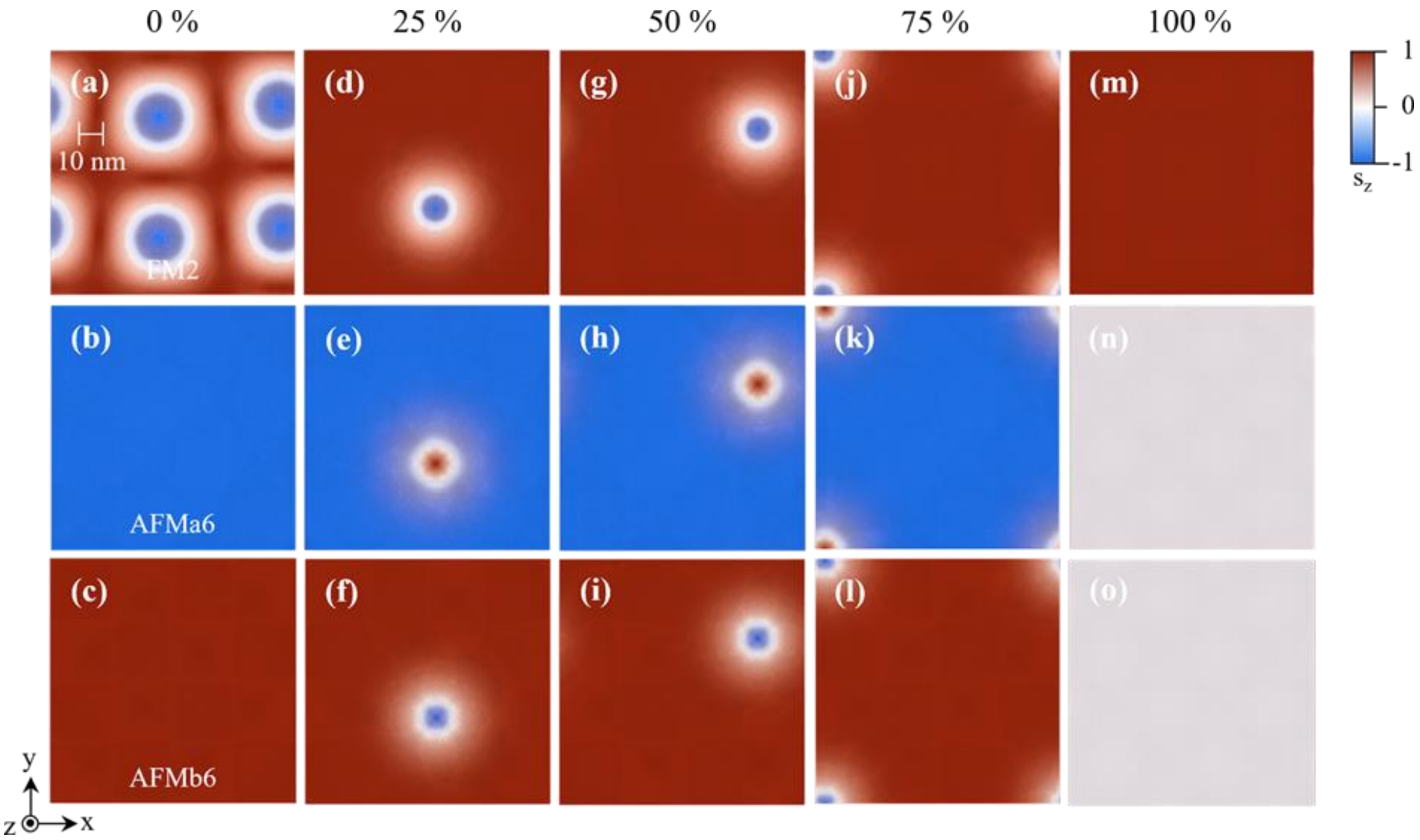

Figure 4. Top view cross section snapshots of the evolution of the final state of the Sk in the FM (first row, $2^{nd}$ ML from the interface), and the AFM in the core (second row: AFMa, third row: AFMb, $6^{th}$ ML from the interface) under the effect of the FM population of NM sites at the interface.

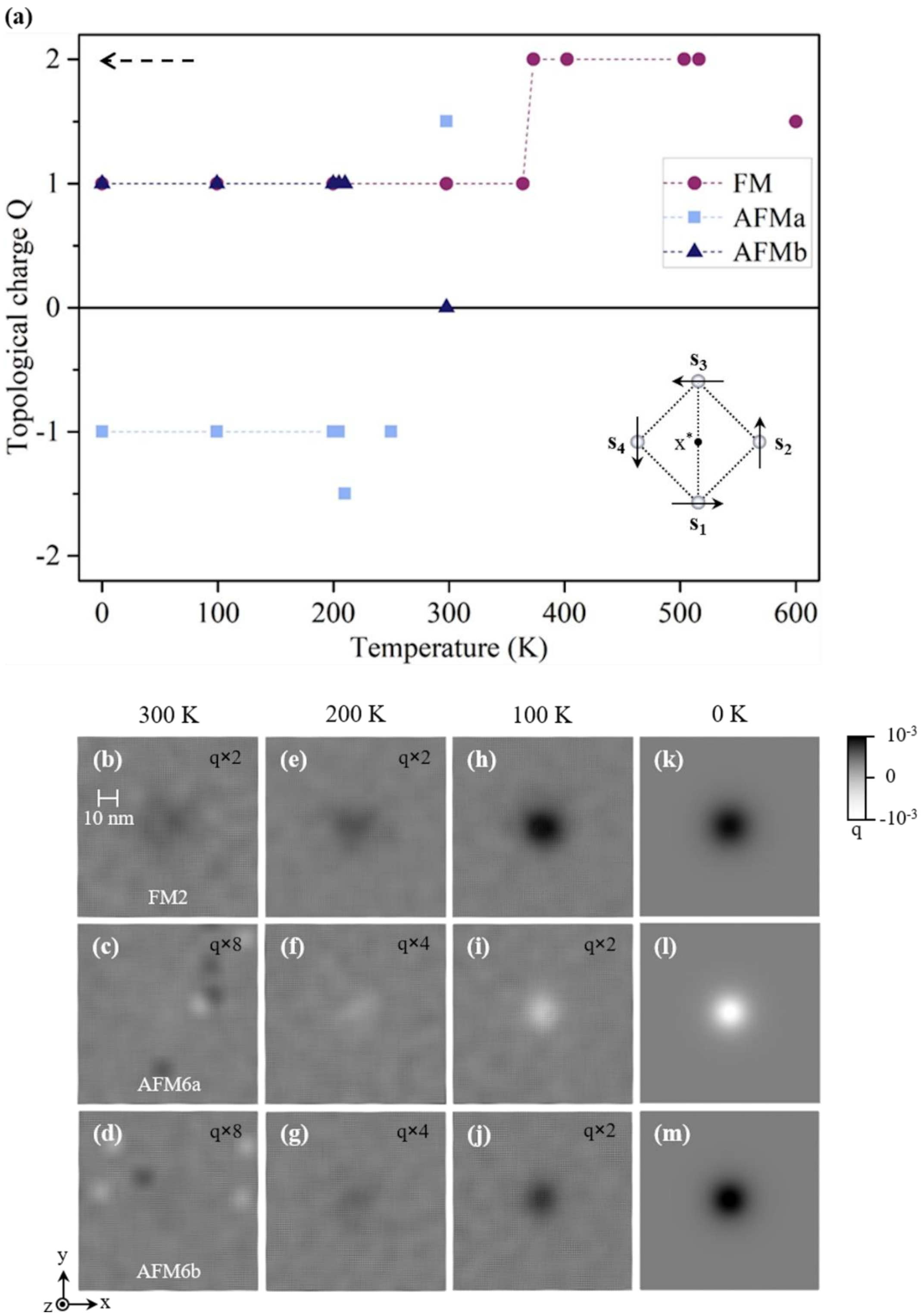

Figure 5. (a) Temperature dependence of the topological charge Q= $\sum_{x^*} q(x^*)$ for the FM and both sublattices of the AFM. The dotted arrow indicates the timeline. Insert: schematic showing the discretization method for calculating q, for a two-dimensional square cell. (b-m) Spatial distribution of the topological charge density q after post-processing the raw data images [Fig. S10] with a Gaussian filter for better visibility. Note that Q was calculated from the unprocessed q.

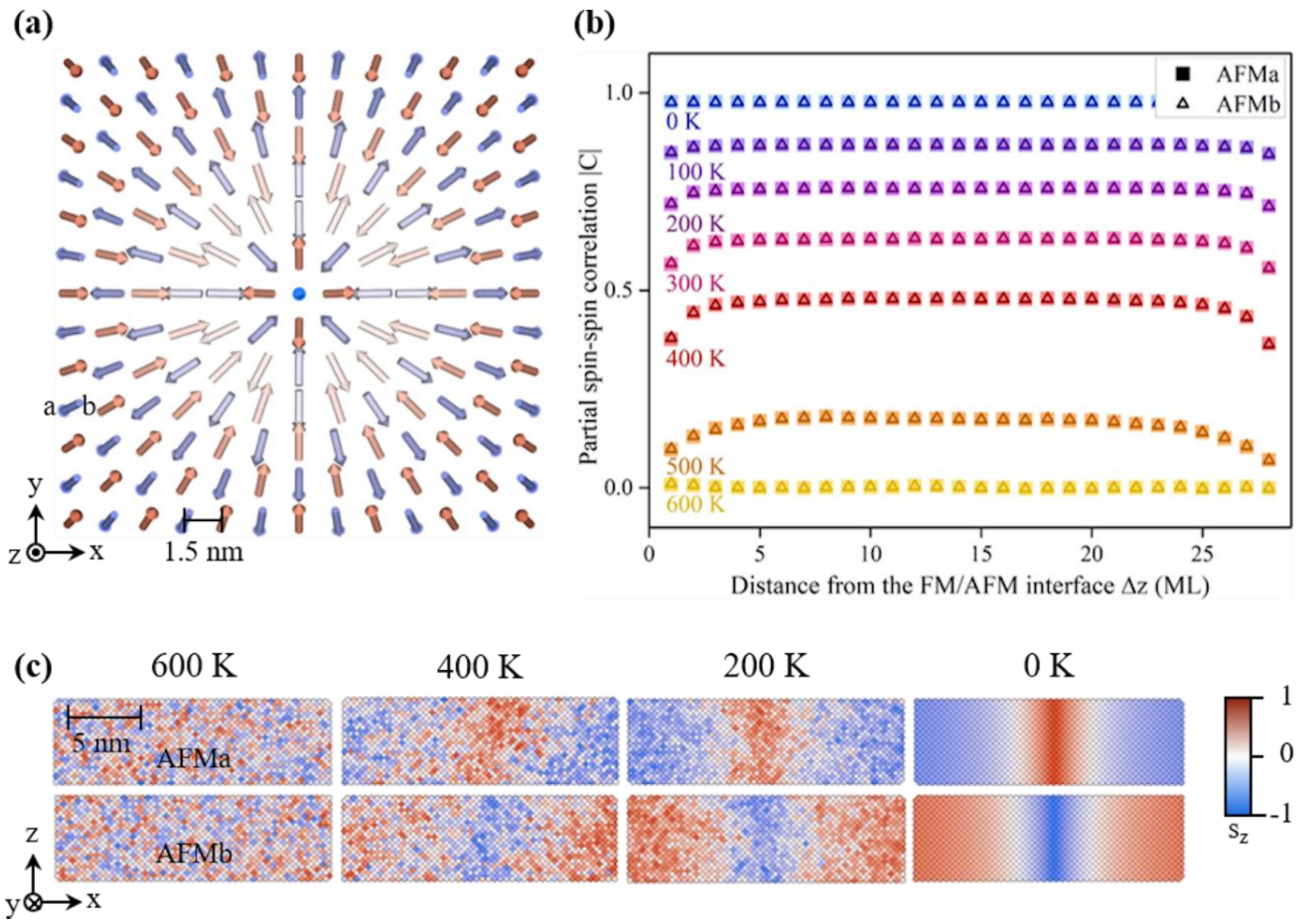

Figure 6. (a) Zoom on the top view cross section of the intertwined AFMa- and AFMb-sublattice Sk, imprinted in the 28 MLs-thick AFM (6th ML from the FM/AFM interface), at remanence, following thermal and magnetic field cycling procedures. To facilitate the reading of the image, only 1 out of every 7 spins is shown. (b) Temperature-dependence of the modulus of the sublattice-resolved partial spin-spin correlation function applied in the center of the spin texture imprinted in the AFM. (c) Side view cross section at selected temperatures, showing the formation of the Sk tube in the AFM.

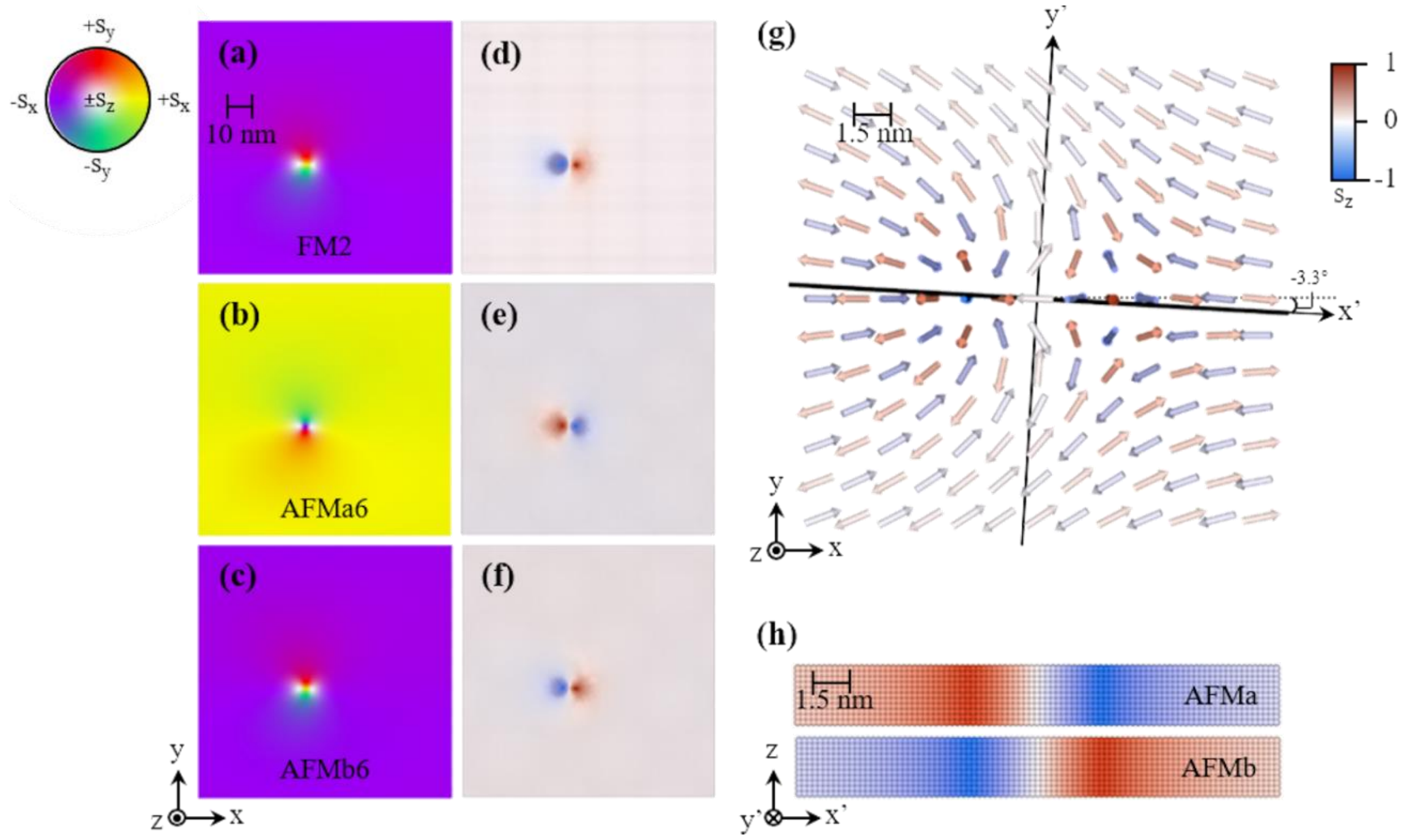

Figure 7. Top view cross section snapshots of the final state of the Bm in the FM (first row, 2nd ML from the FM/AFM interface), and the AFM in the core (second row: AFMa sublattice, third row: AFMb sublattice, 6th ML from the FM/AFM interface) showing (a-c) the in-plane and (d-f) out-of-plane component of the spins. (g) Zoom on the top view cross section of the intertwined AFMa- and AFMb-sublattice Bm, imprinted in the 11 MLs-thick AFM (6th ML from the FM/AFM interface), at remanence, following thermal and magnetic field cycling procedures. To facilitate the reading of the image, only 1 out of every 7 spins is shown. (h) Side view cross section, showing the Bm tube in the AFM, after the imprinting process.

## Supplemental material:

# Imprinting of skyrmions and bimerons in an antiferromagnet

Coline Thevenard,[1] Miina Leiviskä,[1] Richard F. L. Evans,[2] Daria Gusakova,[1] Vincent Baltz[1]

[1]*Univ. Grenoble Alpes, CNRS, CEA, Grenoble INP, IRIG-SPINTEC, F-38000 Grenoble, France*

[2]*School of Physics, Engineering and Technology, University of York, York YO10 5DD, United Kingdom*

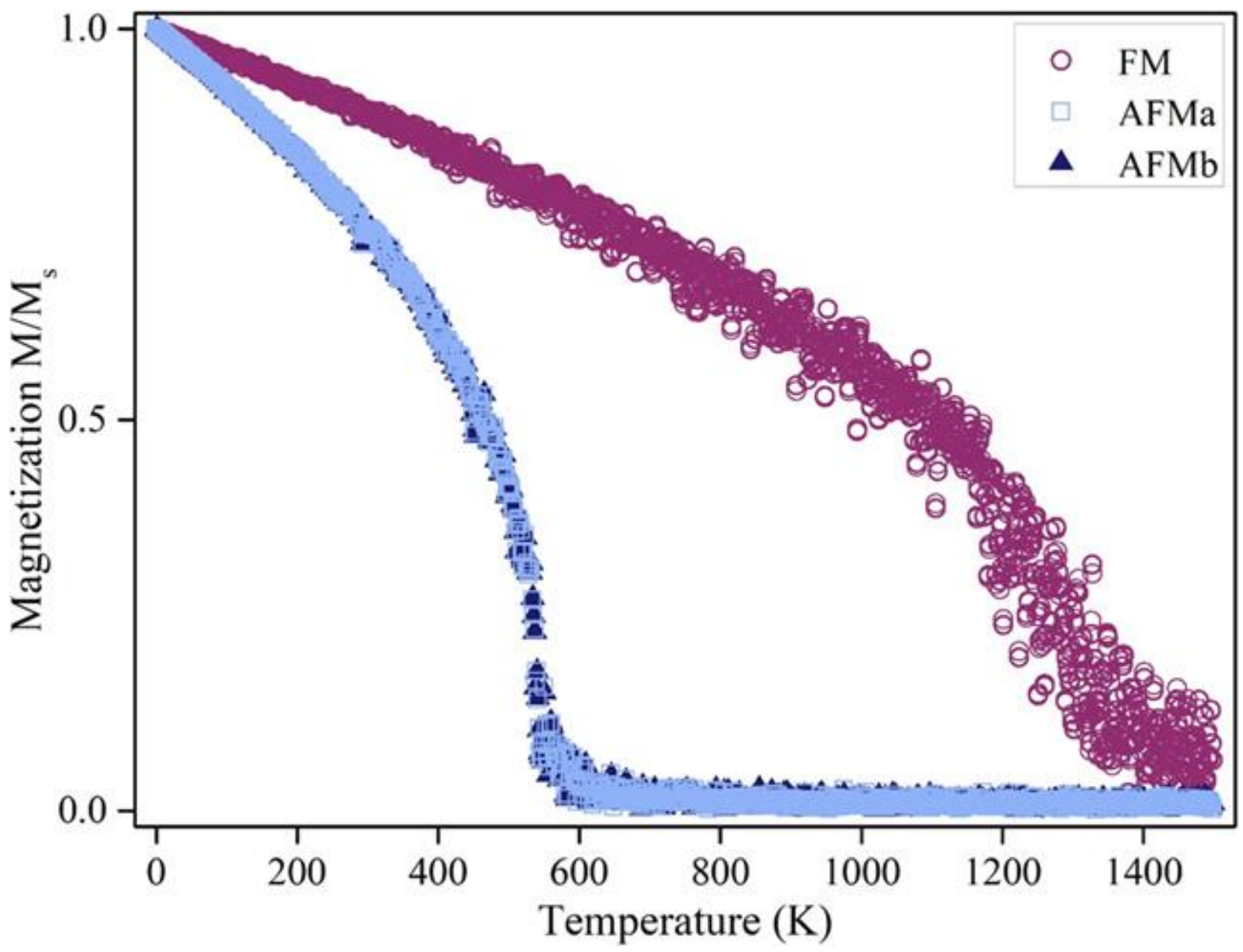

Figure S1: Curie ($T_C$) and Néel ($T_N$) temperatures of the FM and AFM respectively, obtained via simulations of the temperature dependence of the normalized magnetization ($M/M_s$) of each magnetic lattice and sublattice in the stack.

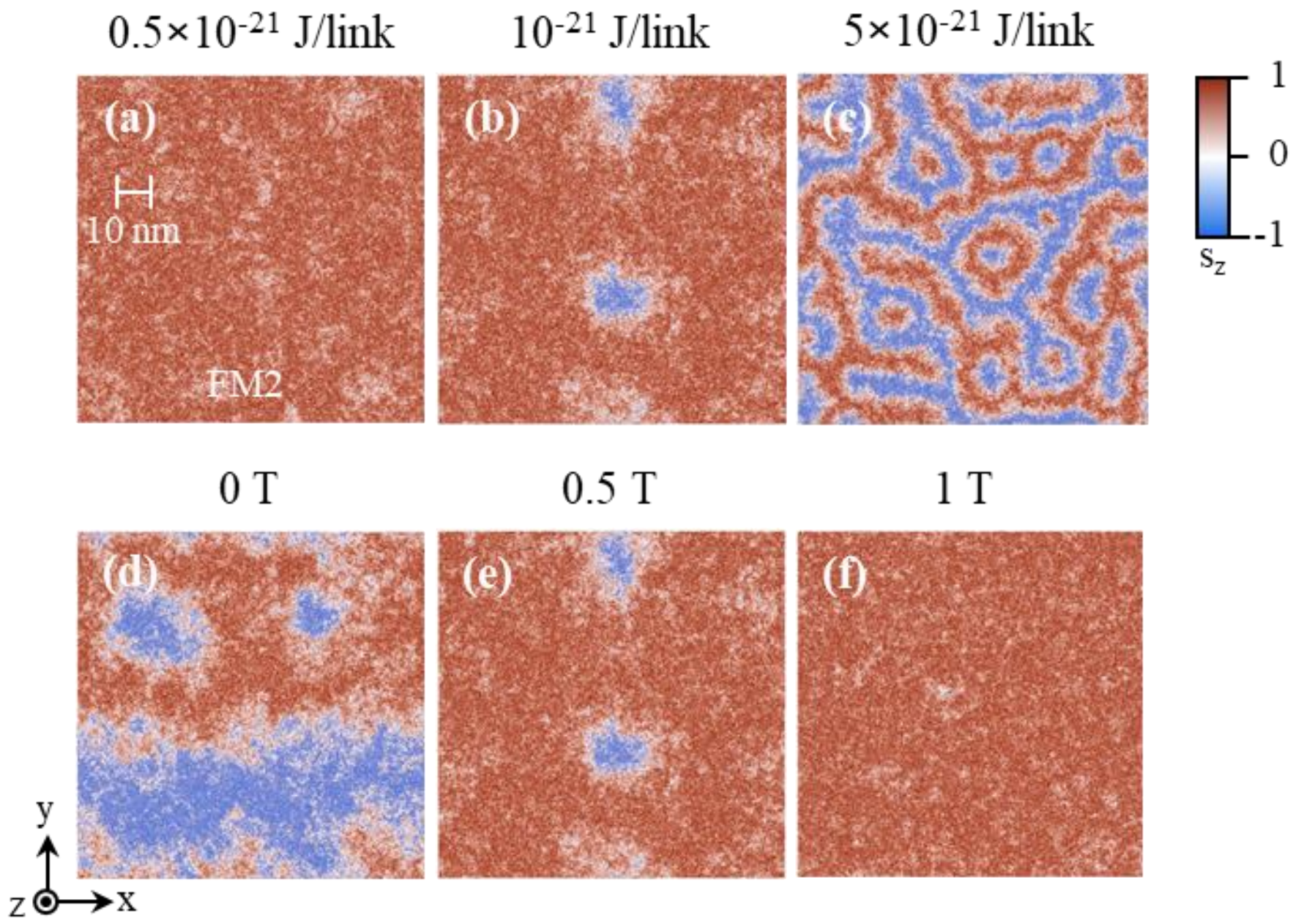

Figure S2: Top view cross section snapshots, at 600 K, of the evolution of the spin textures in the FM (2nd ML from the interface) under the effect of (a-c) the DMI parameter under an applied magnetic field of 0.5 T and (d-f) the strength of the applied magnetic field under a DMI parameter of $10^{-21}$ J/link. All other parameters are the same as in the main text, for the case of out-of-plane anisotropy.

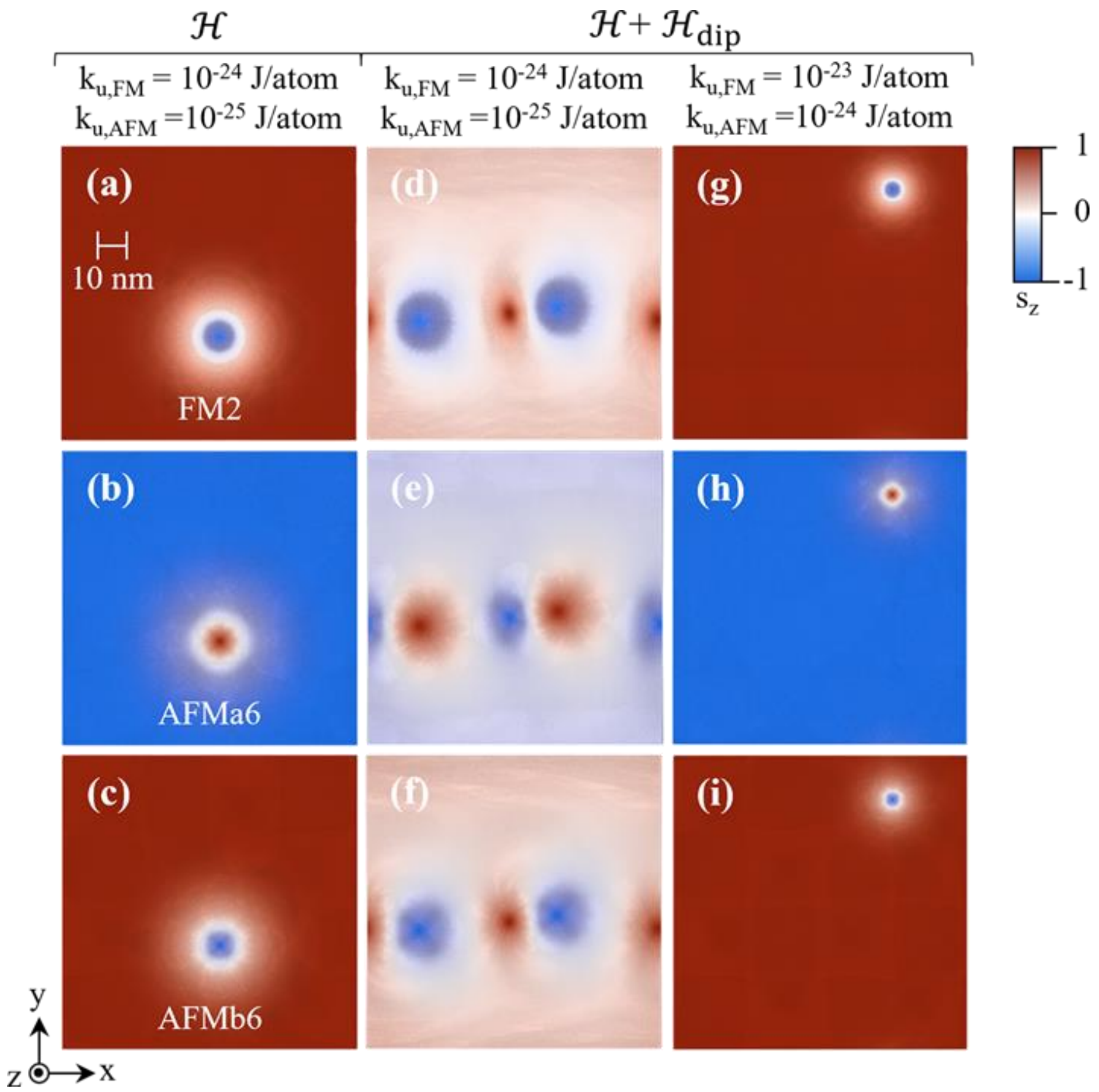

Figure S3: Top view cross section snapshots of the evolution of the final state of the Sk in the FM (first row, 2nd ML from the interface), and the AFM in the core (second row: AFMa, third row: AFMb, 6th ML from the interface) under the effect of the dipolar interaction and of the uniaxial anisotropy of FM and AFM.

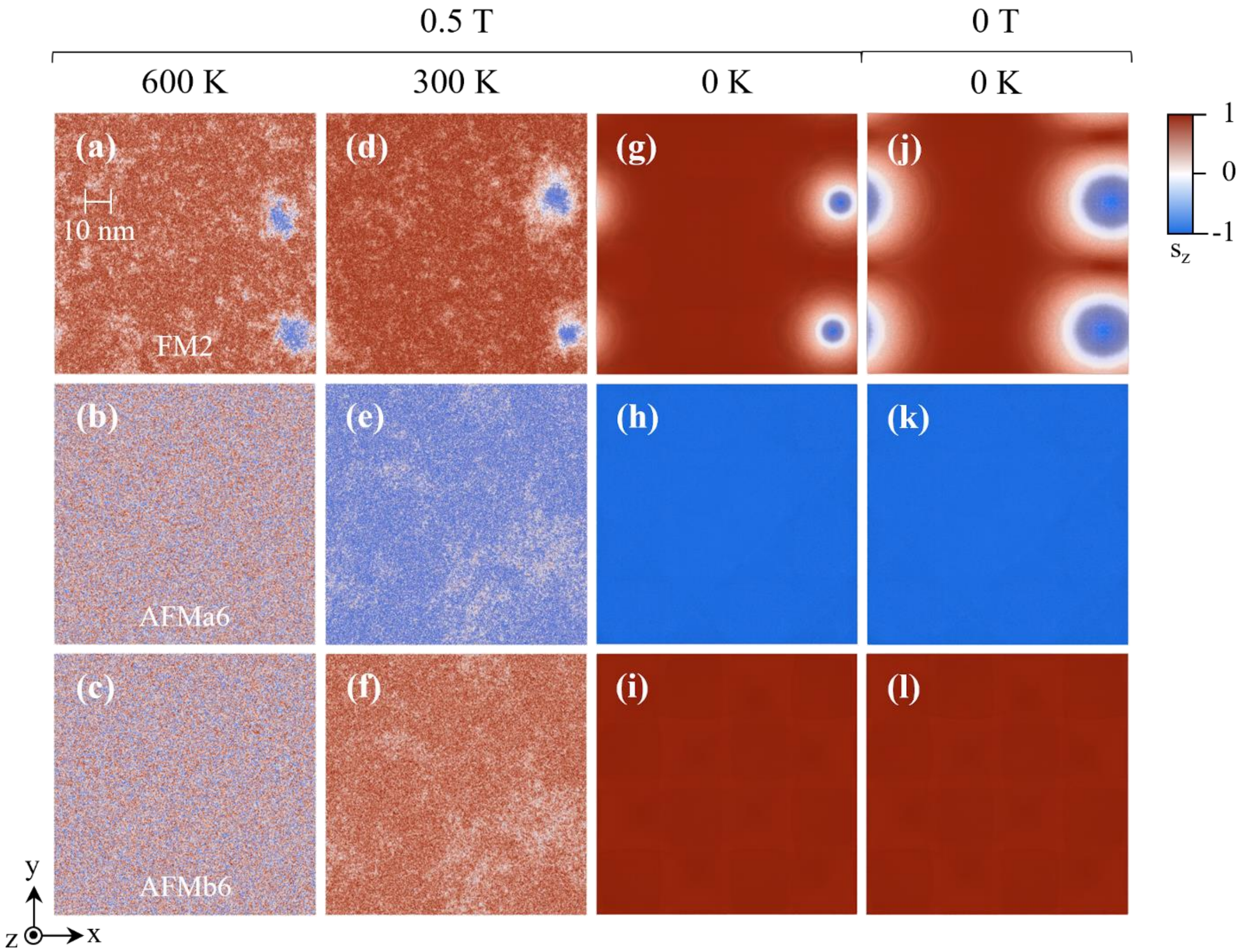

Figure S4. Top view cross section snapshots of the evolution of the Sk in the FM (first row, 2nd ML from the FM/AFM interface), and the AFM in the core (second row: AFMa sublattice, third row: AFMb sublattice, 6th ML from the FM/AFM interface) for $J_{FM-AFM}$ = 0 J/link. First column: field-induced nucleation in the FM, below $T_C$ and above $T_N$. Second and third columns: imprinting and stabilization of Sks in the AFM, via cooling across $T_N$. Fourth column: remanence state.

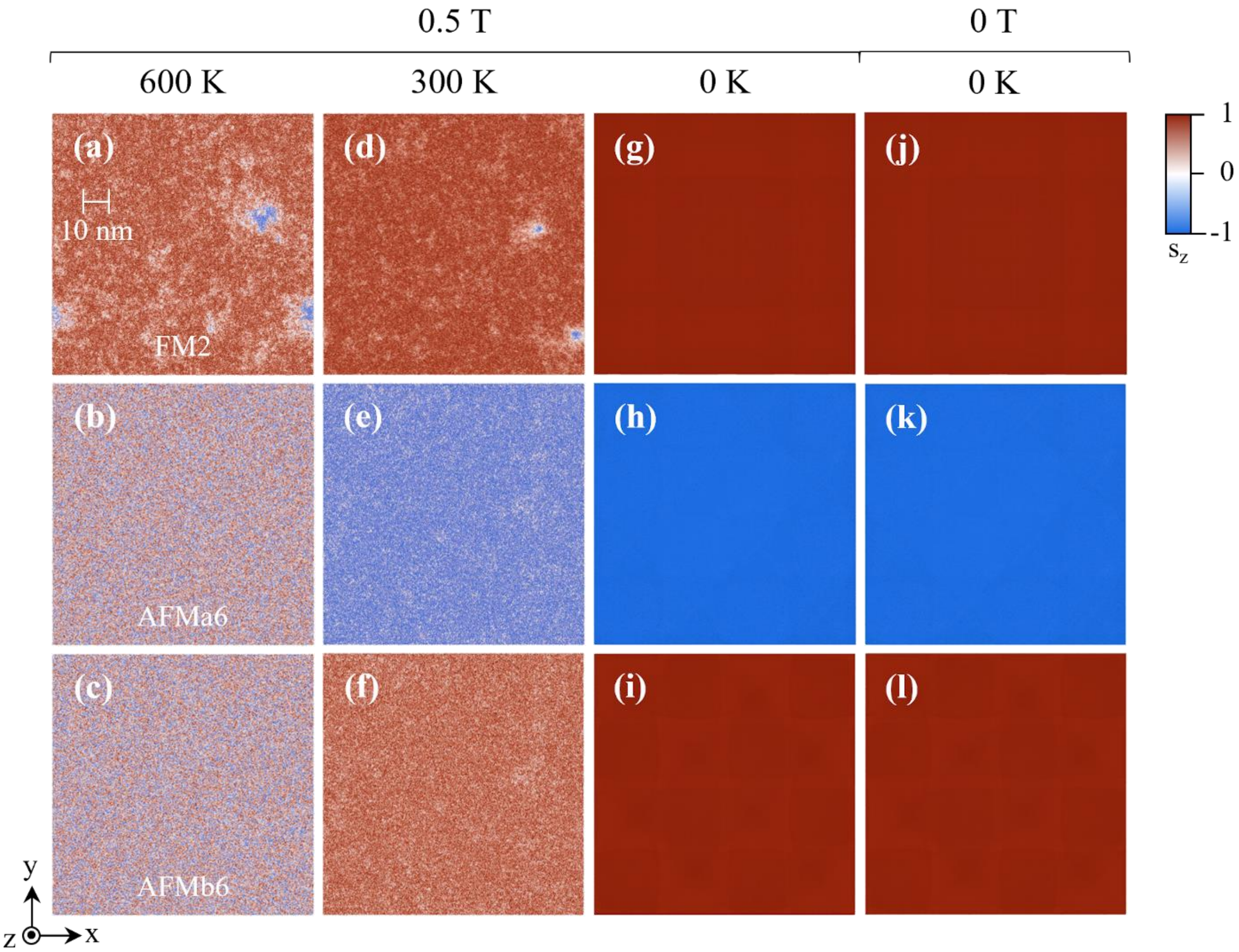

Figure S5. Top view cross section snapshots of the evolution of the Sk in the FM (first row, 2nd ML from the FM/AFM interface), and the AFM in the core (second row: AFMa sublattice, third row: AFMb sublattice, 6th ML from the FM/AFM interface) for $J_{FM-AFM}$ = 5×10$^{-22}$ J/link. First column: field-induced nucleation in the FM, below $T_C$ and above $T_N$. Second and third columns: imprinting and stabilization of Sks in the AFM, via cooling across $T_N$. Fourth column: remanence state.

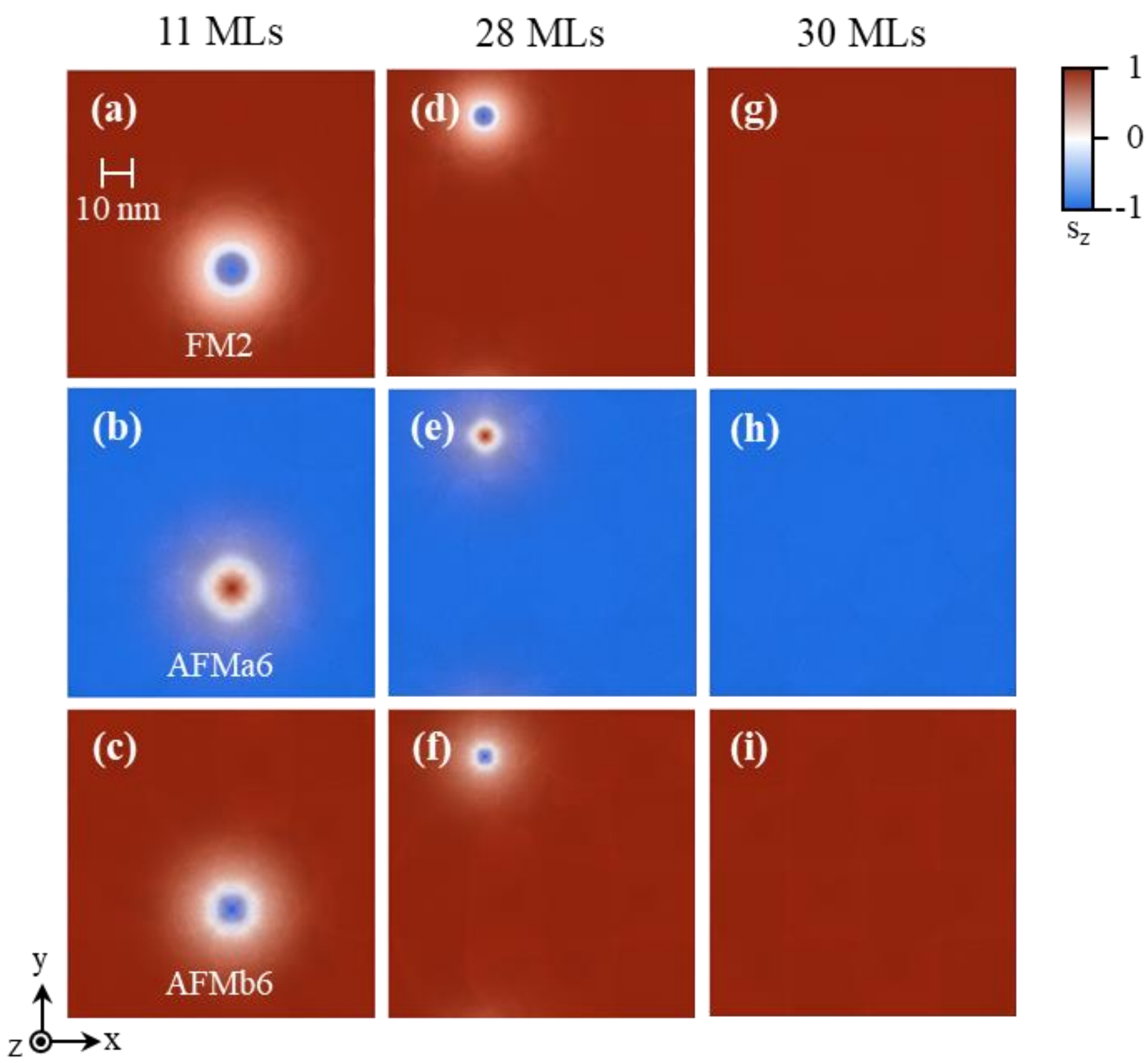

Figure S6: Top view cross section snapshots of the evolution of the final state of the Sk in the FM (first row, 2nd ML from the interface), and the AFM in the core (second row: AFMa, third row: AFMb, 6th ML from the interface) under the effect of the AFM thickness.

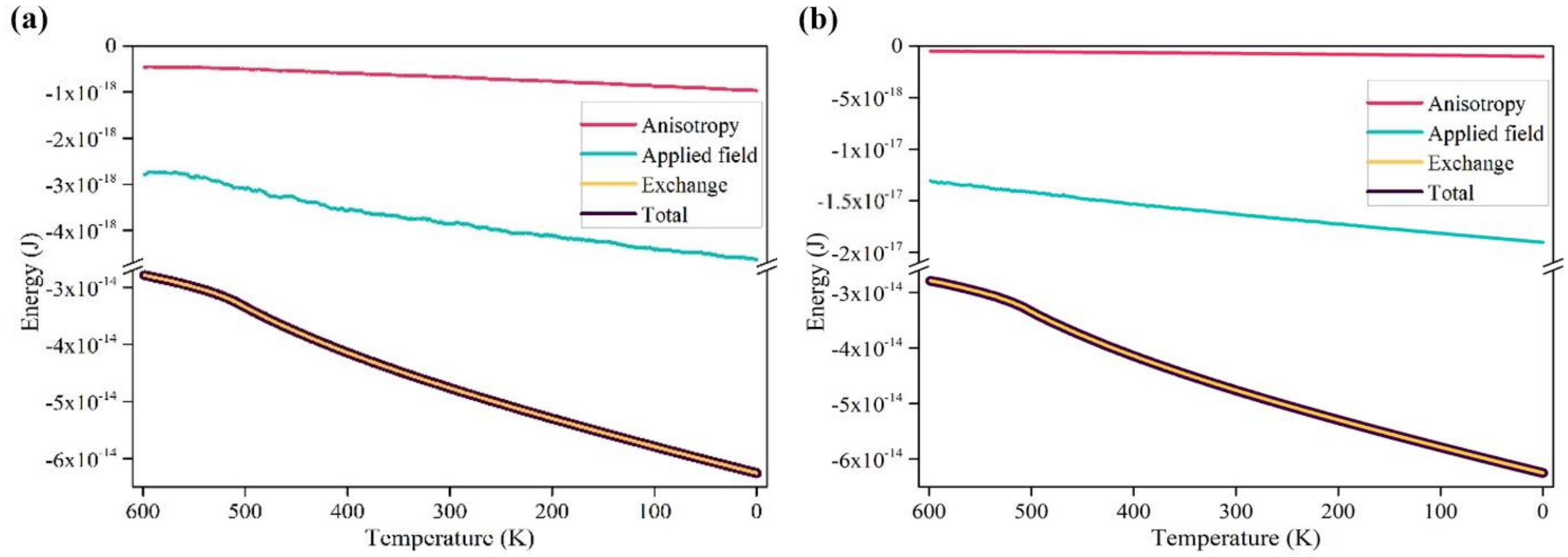

Figure S7: Temperature dependence of the anisotropy, applied field, exchange and total energy for B = (a) 0.5 T with Sk in the FM and AFM and (b) 2 T without Sk. Time dependence can be derived from $T[t] = T_0 \exp\left(-\frac{t}{\tau}\right)$ with $T_0$ = 600 K the initial temperature and $\tau$ = 0.4 ns the cooling time.

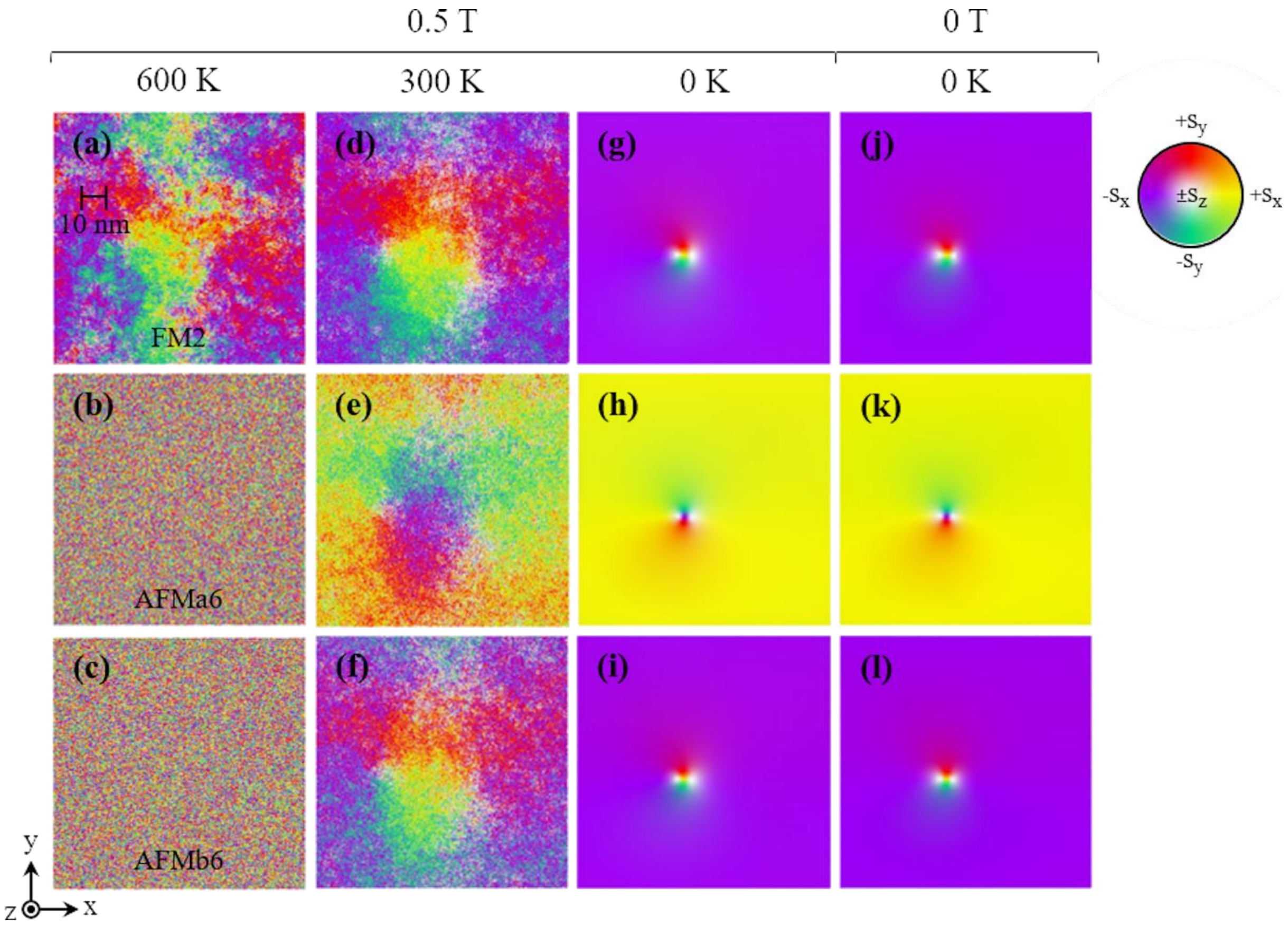

Figure S8: Top view cross section snapshots of the evolution of the in-plane components of the Bm in the FM (first row, 2nd ML from the interface), and the AFM in the core (second row: AFMa, third row: AFMb, 6th ML from the interface). First column: field-induced nucleation in the FM, below $T_C$ and above $T_N$. Second and third columns: imprinting and stabilization of the Bm in the AFM, via cooling across $T_N$. Fourth column: remanence state.

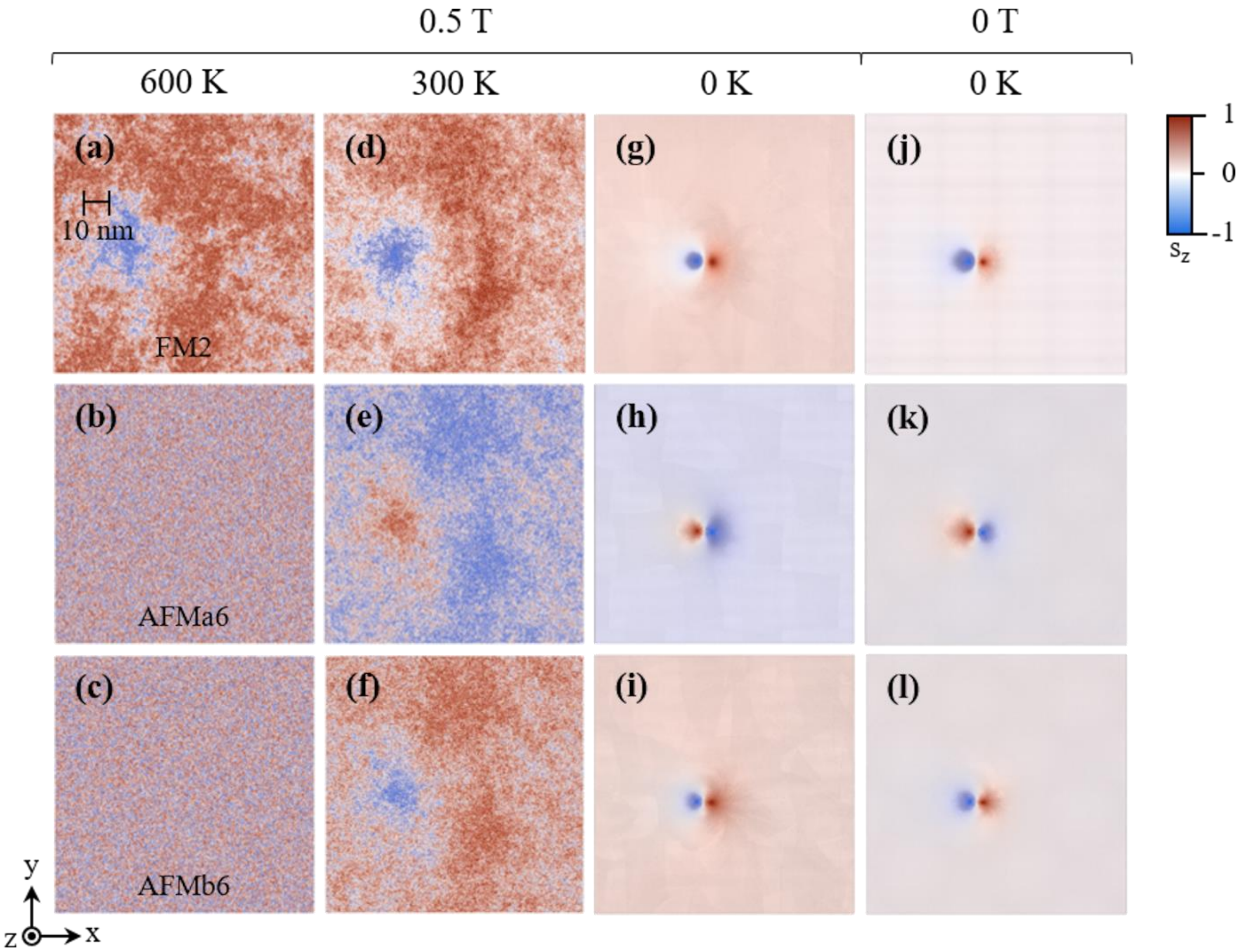

Figure S9: Same as Fig. S8 for the out-of-plane components.

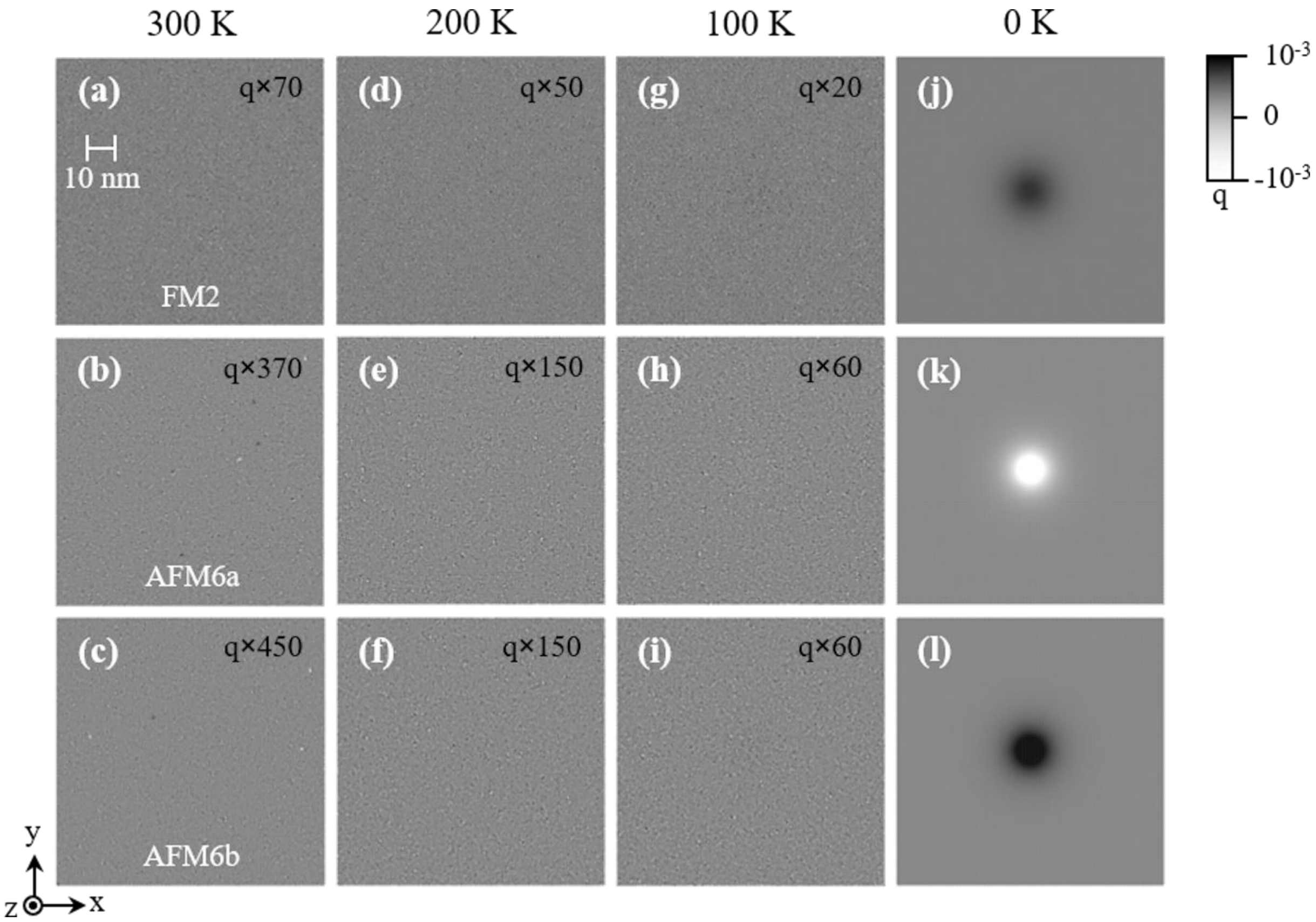

Figure S10: Spatial distribution of the topological charge density (q) in the FM (2nd ML from the interface) and the 11 MLs-thick AFM in the core (6th ML from the interface), for each sublattice: AFMa and AFMb, obtained from the raw data. Note that Q was calculated from these unprocessed q. To obtain Fig. 5(b-m) in the main text these raw data were post-processed with a Gaussian filter of the form: $\frac{\exp\left(-\frac{x^2+y^2}{2\sigma^2}\right)}{\sum_{(\Delta x,\Delta y)\in\left[-\frac{a(n-1)}{4},\frac{a(n-1)}{4}\right]^2}\exp\left(-\frac{(x+\Delta x)^2+(y+\Delta y)^2}{2\sigma^2}\right)}$, with a = 4.17 Å, n=17 and $\sigma = \sqrt{20}$. This image post-processing makes the tiny spots in q visible.

| Parameters | Simulation units | | SI units | | Simulation to SI units conversion formula |
|---|---|---|---|---|---|
| $\mathbf{J_{FM}}$ | $13.3\times10^{-21}$ | J/link | $6.38\times10^{-11}$ | J/m | 2/a |
| $\mathbf{J_{AFM}}$ | $-5.04\times10^{-21}$ | J/link | $-2.42\times10^{-11}$ | J/m | 2/a |
| $\mathbf{J_{FM-AFM}}$ | $10^{-21}$ | J/link | $4.80\times10^{-12}$ | J/m | 2/a |
| $\mathbf{D_{HM-FM}}$ | $10^{-21}$ | J/link | $2.30\times10^{-2}$ | J/m² | $(2/a)^2$ |
| $\mathbf{k_{N,HM-FM}}$ | $2\times10^{-24}$ | J/atom | $4.60\times10^{-5}$ | J/m² | $n_s\ (2/a)^2$ |
| $\mathbf{k_{u,FM}}\ (Sk)$ | $10^{-24}$ | J/atom | $1.10\times10^{5}$ | J/m³ | $n_{at}\ (2/a)^3$ |
| $\mathbf{k_{u,FM}}\ (Bm)$ | $-10^{-23}$ | J/atom | $-1.10\times10^{6}$ | J/m³ | |
| $\mathbf{k_{u,AFM}}\ (Sk)$ | $10^{-25}$ | J/atom | $1.10\times10^{4}$ | J/m³ | |
| $\mathbf{k_{u,AFM}}\ (Bm)$ | $-10^{-24}$ | J/atom | $-1.10\times10^{5}$ | J/m³ | |

Table S1: Conversion between simulation and SI units. We recall that the crystal structure is consistently set to that of two interpenetrating single cubic lattices and equivalent to rocksalt, with a lattice parameter a = 4.17 Å, corresponding to a unit cell size a/2 and an equivalent number of atoms per unit cell $n_{at} = 1$ or per unit surface $n_s = 1$.